\documentclass[twoside,a4paper,11pt]{sca}
\usepackage{graphicx}
\usepackage{hyperref}
\usepackage{movie15}
\usepackage{natbib}  
\begin{document}
\pagenumbering{arabic}
\pagestyle{myheadings}
\thispagestyle{empty}
{\flushright\includegraphics[width=\textwidth,bb=90 650 520 700]{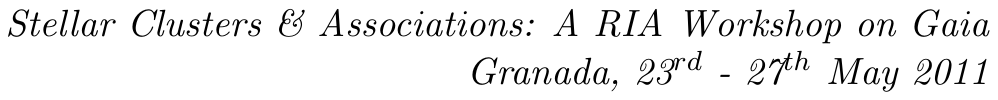}}
\vspace*{0.2cm}
\begin{flushleft}
{\bf {\LARGE
%
The Star Cluster Populations of Compact Galaxy Groups.
%
}\\
\vspace*{1cm}
%
I.~S.~Konstantopoulos$^{1}$,
K.~Fedotov$^{2}$,
S.~C. Gallagher$^{2}$,
A.~Maybhate$^{3}$,
P.~R.~Durrell$^{4}$,
J.~C. Charlton$^{1}$
%
}\\
\vspace*{0.5cm}
%
$^{1}$
The Pennsylvania State University, 525 Davey Lab, 
University Park, PA 16802, USA\\
$^{2}$
The University of Western Ontario, London, ON, N6A 3K7, Canada\\
$^{3}$
Space Telescope Science Institute, Baltimore, MD\\
$^{4}$
Youngstown State University, Youngstown, OH 44555
%
\end{flushleft}
%
\markboth{
Star Clusters in Galaxy Groups
}{ 
%
Iraklis S. Konstantopoulos
%
}
\thispagestyle{empty}
\vspace*{0.4cm}
\begin{minipage}[l]{0.09\textwidth}
\ 
\end{minipage}
\begin{minipage}[r]{0.9\textwidth}
\vspace{1cm}
\section*{Abstract}{\small
%
Star clusters are ideal tracers of star formation activity 
in systems outside the volume that can be studied using 
individual, resolved stars. These unresolved clusters span
orders of magnitude in brightness and mass, and their formation
is linked to the overall star formation in their host galaxy. 
In that sense, the age distribution of a cluster population 
is a good proxy of the overall star formation history of the 
host. 

This talk presents a comparative study of clusters in seven
compact galaxy groups. The aim is to use the cluster age 
distributions to infer the star formation history of these 
groups and link these to a proposed evolutionary sequence
for compact galaxy groups. 
%
\normalsize}
\end{minipage}
%
%
%
\section{Introduction}\label{sec:intro}
Compact galaxy groups (CGs) occupy an interesting part of the
parameter space that pertains to the clumping of matter in the
universe. On the one hand, they lie at the low end tail of the 
distribution of membership size, as they contain few galaxies -- 
typically three or four \citep{hickson82}. On the other hand, 
their compactness places them on the high end of the number 
density distribution, similar to that in the centres of galaxy
clusters \citep{dressler80,brad90}. 

In addition, CGs display low velocity dispersions of 
$\sigma_\textup{\scriptsize CG}\sim250~$km$\,$s$^{1}$ 
\citep{tago08,asqu}, compared to galaxy cluster dispersions of 
$\sigma_\textup{\scriptsize cluster}\sim750~$km$\,$s$^{1}$ 
\citep{binggeli87,the86,asqu}. This dynamical situation 
lengthens interactions, when such events occur, or forces galaxies
to a state of quasi-secular evolution: while they may not interact 
physically with their neighbours, they are always affected
by them dynamically \citep{martig08,isk10}. In terms of galaxy
evolution, they present potential precursors of isolated 
ellipticals. Furthermore, they allow for close studies of galaxy 
evolution and morphological transformation, given their relatively 
few degrees of freedom (as compared with galaxy clusters and their 
hundreds of members). 

\begin{figure}[!t]
\center
\includegraphics[width=0.8\textwidth, angle=270]{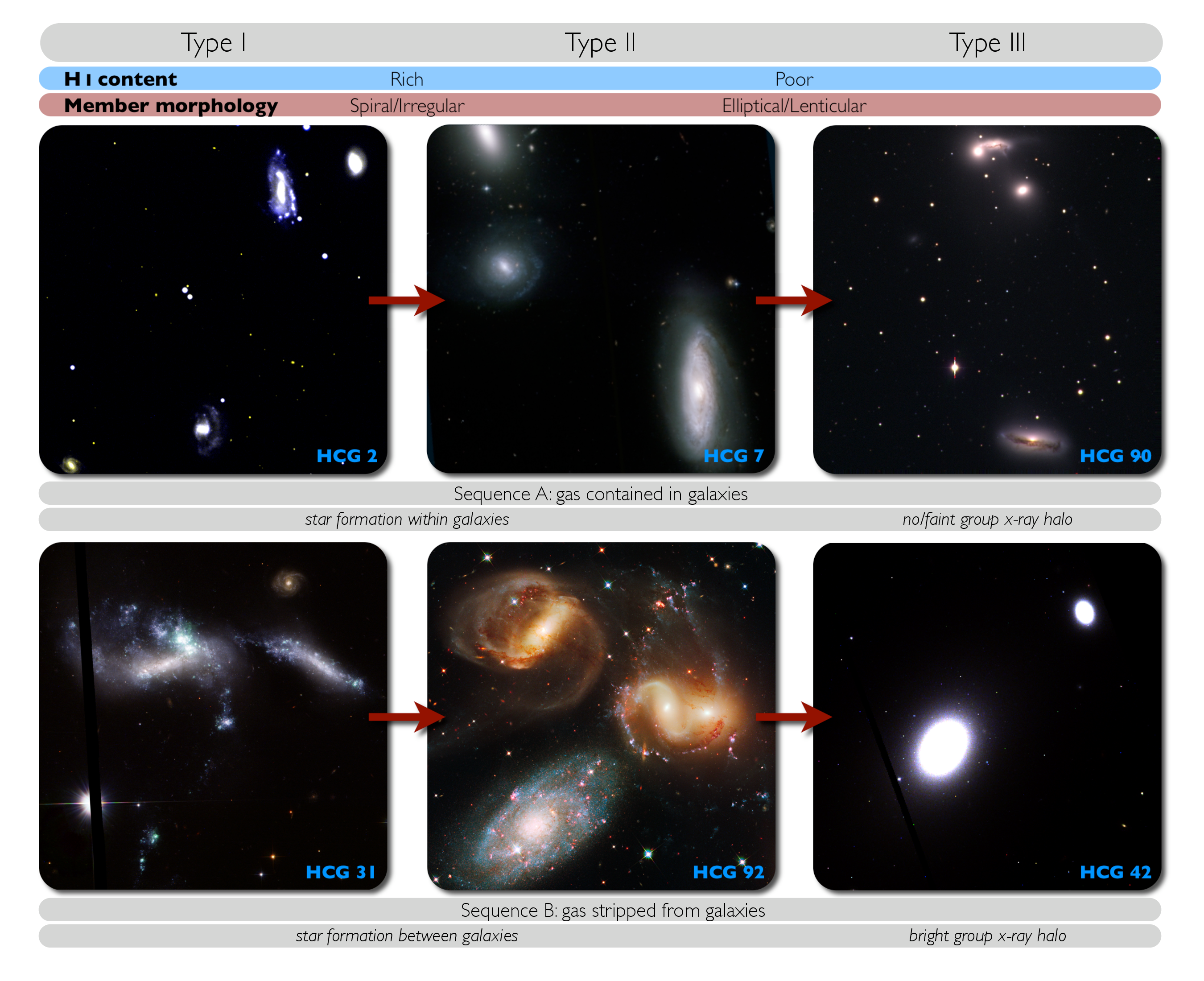}
\caption{
The evolutionary diagram proposed by \citet{isk10}. Gas content
decreases from left to right, according to the definition of 
\citet{johnson07}. The top sequence includes groups that hold the 
H\,{\sc i} gas within the member galaxies, while the CGs of the 
bottom sequence have developed a gaseous (and stellar) intra-group 
medium through physical interactions. 
The top sequence includes the possible precursors of dry mergers, 
as galaxies within these groups evolve separately and are likely 
to have a very low, or no gas content when they eventually merge. 
}\label{fig:evo}
\end{figure}

The work presented here relates qualitative age dating of thousands 
of star clusters in seven Hickson compact groups (HCGs) to the 
CG evolutionary diagram proposed by \citet[Fig.~\ref{fig:evo}]{isk10}. 
This draws upon results presented in an ongoing series of papers on 
individual groups 
\citep{gallagher10,isk10,fedotov11}, 
and on the collective properties of HCGs 
\citep{johnson07,gallagher08,walker10,tzanavaris10}. 
In brief, the diagram classifies galaxy groups according to a
gas richness criterion. With the basic assumption that H\,{\sc i}
represents the reservoir of gas available for further star formation
(star formation potential), \citet{johnson07} uses the ratio of 
gas mass to dynamical mass (total mass) to split CGs into three
types: I, II, and III, for gas-rich, intermediate and gas-poor. 
This is then filtered through the spatial distribution of the 
H\,{\sc i} gas \citep[cf.]{vm01}, to give rise to a two-pronged
diagram, shown in Figure~\ref{fig:evo}. One sequence maps CGs where 
the gas in contained wholly within the galaxies; the other includes 
groups that have started to build an intra-group medium through the 
release of gas during interactions. 


\section{Methodology}\label{sec:methodology}
We study star clusters in seven Hickson CGs (HCGs) following 
the methodology presented in \citet{gallagher10}. This is refined
by using more recent simple stellar population (SSP) models by
\citet{marigo08}. We compare the \textit{HST}-ACS $B-V$ (F435W--F606W)
and $V-I$ (F606W--F814W) colours of the clusters to these SSP models
and infer their ages. Lacking photometric coverage in the $U$-band, 
we cannot break the age-extinction degeneracy inherent in the 
$BVI$ baseline. We therefore compare CG cluster ages qualitatively. 

We are largely aided by the fact that CGs are deficient in H\,{\sc i}
gas \citep{haynes84}: since neutral gas is related to dust 
\citep[e.\,g.][]{pohlen10}, this implies an overall low dust content 
for HCGs, thus allowing for a `clearer', relatively unextinguished 
view of their cluster populations. 
In addition, we take advantage of the long bandpass
of the F606W filter, which covers the H$\alpha$ line, by synthesising
a Starburst99 \citep{sb99} SSP model that includes emission lines
(i.\,e. from H$\beta$, [O{\sc iii}], H$\alpha$ and [N{\sc ii}]). 
In the observations, an excess in $V$-band light is interpreted as
nebular emission arising from the ionisation of residual gas from
star formation around a young cluster. Sources with such an excess 
are considered to be no older than 10~Myr, a timescale appropriate
for the dispersal of gas from the maturing generation of supernov\ae. 

\section{Results}\label{sec:results}
\begin{figure}[!h]
\center
\includegraphics[width=0.5\textwidth]{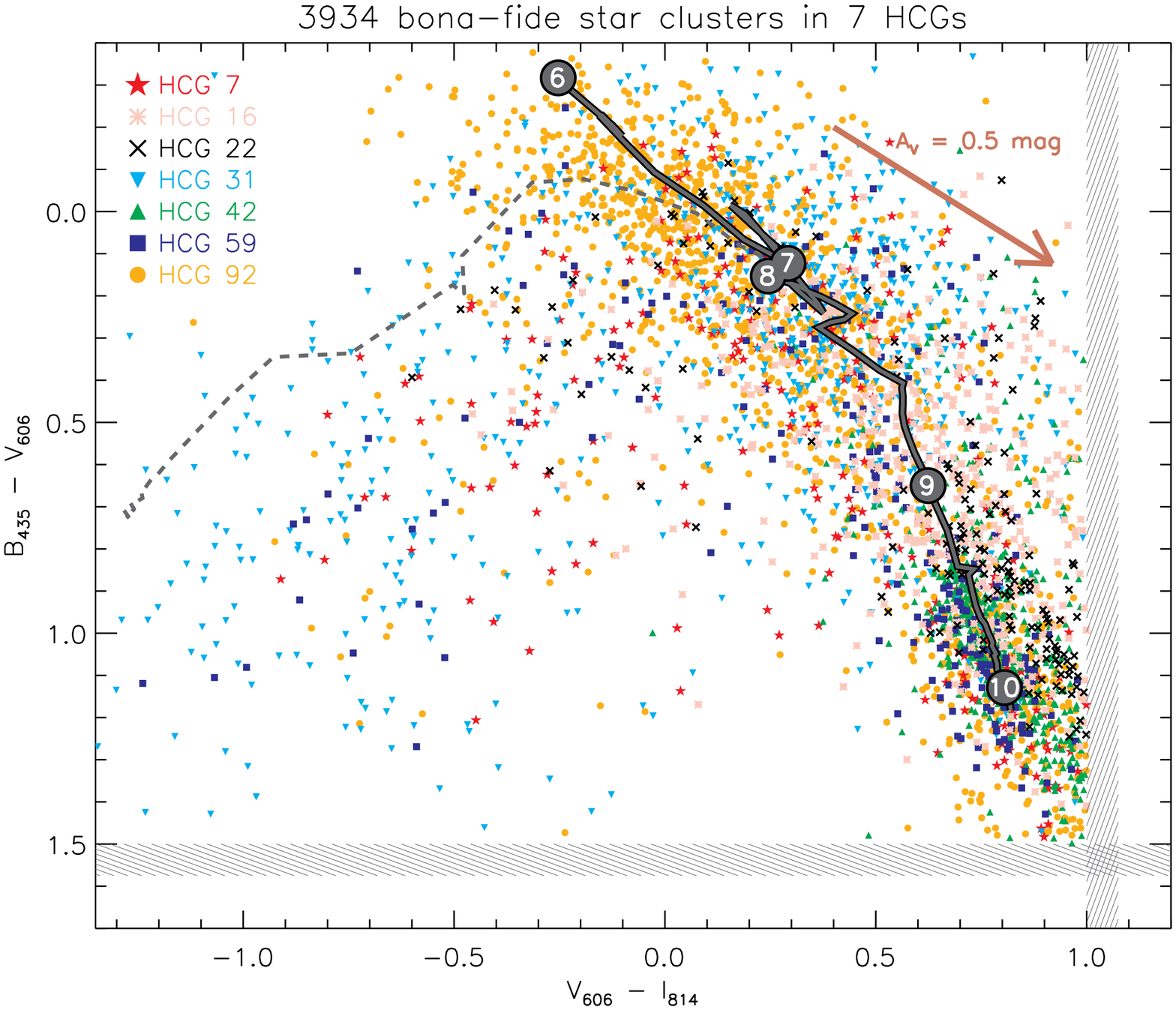}~\includegraphics[width=0.5\textwidth]{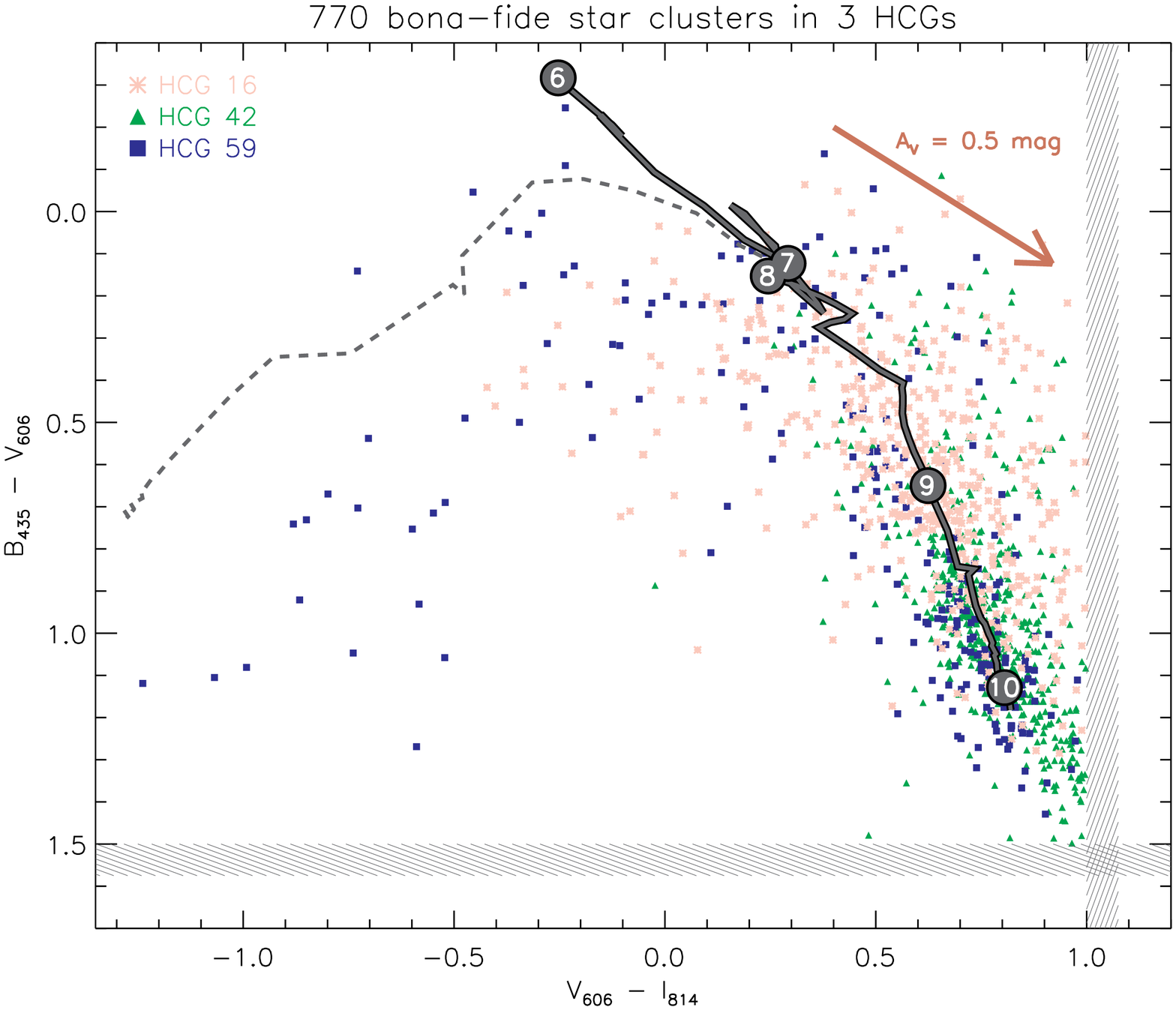}
\caption{
{\bf Left:} 
The star cluster populations of seven compact galaxy 
groups. We only consider clusters brighter than $M_V=-9$~mag, in 
order to eliminate contamination from giant stars local to the 
host galaxies. Furthermore, a selection based on PSF-photometry 
ensures little or no contamination from foreground stars and 
background galaxies. 
The solid line plots a \citet{marigo08} SSP model, while the dashed 
line represents a Starburst99 model \citep{sb99} that includes the
emission lines transmitted in the $V$ (F606W) band. Thus, 
any source shifted to `greener' values is considered to be younger 
than 10~Myr old, a timescale appropriate for the clearing of the 
natal gas of clusters from supernov\ae.
\newline%
The evolutionary state of a group is closely linked with the cluster
age distribution: star forming groups show a large concentration 
of `nebular' sources (those bracketed by the dashed line) and many 
bright, blue clusters (ages less than a few hundred Myr); groups
containing both star forming and evolved/quiescent galaxies show 
an even spread of nebular, young and old (red) sources; while 
evolved groups, those containing mostly or entirely quiescent 
galaxies, show few clusters outside the tight, red globular cluster 
clump -- centred around \mbox{($V-I$, $B-V$)} $\simeq$ (0.8, 1.1). 
\newline
%
Continuing along that line, the {\bf right panel} selects groups 
along the proposed evolutionary sequence of \citet{isk10}, where 
HCGs~16, 59, and 42 represent classes I, II, and III, as defined 
in \citet{johnson07}. The correlations between the mentioned 
sequence and the cluster populations are discussed in detail in 
Section~\ref{sec:results}. 
}\label{fig:clusters}
\end{figure}

The colour plots of Figure~\ref{fig:clusters} summarise
our results. Groups of different evolutionary stage
feature strikingly different cluster populations (left). 
Selecting according to the 
proposed evolutionary diagram of \citet{isk10} reveals a pattern 
(right panel): the median cluster age relates well to the 
\citet{johnson07} type, therefore echoing the evolutionary sequence 
among those three types, as described in \citet{isk10}. 

More specifically: the `early-type' HCG~16 shows an even 
star formation history (as evidenced by the cluster formation history)
from the present, all the way to 12 or so Gyr. It has a small 
population of globular clusters, perhaps owing to the apparent lack 
of major mergers in its past (it contains only disk galaxies). 

Next in the sequence, the intermediate type HCG~59 shows increased 
star formation in the current era, as compared to HCG~16. This is 
likely due to recent interactions between its member galaxies, 
which have led to bursts of star formation across the group. This
is supported by a plethora of observational evidence, to be presented
in Konstantopoulos~et~al.~(submitted). 

Finally, the `late-type' HCG~42 contains only bulge-dominated, 
evolved galaxies. Its cluster population mirrors that state, as 
it is heavily skewed toward old ages. We find no evidence of massive
cluster formation over the past few Gyr in this system and no 
nebular sources -- i.\,e. no ongoing formation. 

\section{Summary}\label{sec:summary}
We have used the optical colours, and therefore ages, of thousands 
of \textit{HST}-selected star clusters to infer the properties of 
their host galaxies, members of compact groups. We have linked the
distribution of cluster colours to the evolutionary state of their
hosts and provide evidence in favour of the evolutionary sequence
for CGs proposed by \citet{isk10}. While the lack of $U$-band 
coverage prohibits precision age-dating, the dust deficiency of 
the CG environment facilitates the qualitative age-dating of clusters
with $BVI$ only. The full study, including modelling of the cluster
populations with inferences on star cluster evolution, will be 
presented in a future work. 

%
%
\small  
%
\section*{Acknowledgments}   
%
This work was undertaken as part of the HCG collaboration. In 
addition to the authors, credit is due to the following scientists: 
P.~Tzanavaris,
A.~E.~Zabludoff,
D.~M.~Elmegreen,
K.~E.~Johnson,
C.~Gronwall,
J.~English,
A.~E.~Hornschemeier,
R.~Chandar,
J.~S.~Mulchaey, 
as well as others who have participated in several works mentioned
in this paper. 
%

%

\end{document}